\documentclass[11pt,twoside]{article}

\usepackage{asp2006}
\usepackage{epsf}
\usepackage{lscape}

\newcommand{\pref}{\protect\ref}
\newcommand{\soho}{{\em SOHO}} 
\newcommand{\trace}{{\em TRACE}}
\newcommand{\hinode}{{\em Hinode}}

\begin{document}
\title{What Goes Up Doesn't Necessarily Come Down! \-- Connecting the Dynamics of the Chromosphere and Transition Region with TRACE, Hinode and SUMER}   %%% Fill in title
\author{Scott W. \textsc{McIntosh}\altaffilmark{1}, Bart \textsc{De Pontieu}\altaffilmark{2}}   %%% Fill in author names
%\email{mscott\@ucar.edu, bdp\@lmsal.com}
\altaffiltext{1}{High Altitude Observatory, National Center for Atmospheric Research, PO Box 3000, Boulder, CO 80307, USA}
\altaffiltext{2}{Lockheed Martin Solar and Astrophysics Laboratory, Palo Alto, CA 94304, USA}

\begin{abstract} %%% Abstract to run on from here.
We explore joint observations of the South-East limb made by \hinode{}, \trace{} and \soho/SUMER on April 12, 2008 as part of the Whole Heliosphere Interval (WHI) Quiet Sun Characterization targeted observing program. During the sequence a large, 10Mm long, macro-spicule was sent upward and crossed the line-of-sight of the SUMER slit, an event that affords us an opportunity to study the coupling of cooler chromospheric material to transition region emission formed as hot as 600,000K. This short article provides preliminary results of the data analysis.
\end{abstract}

The Solar Optical Telescope \citep[SOT;][]{Tsuneta2008} of \hinode{} \citep{Hinode} has revolutionized our view of the dynamic chromosphere revealing the existence of at least two types of spicule \citep[][]{DePontieu2007c}. ``Type-I'' spicules are long-lived (3-5 minutes) and exhibit longitudinal motions of the order of 20km/s that are driven by shocks resulting from the leakage of p-modes in regions around strong magnetic flux concentrations \citep[][]{Hansteen2006, DePontieu2007a}. ``Type-II'' spicules, on the other hand, are a relative unknown that show much shorter lifetimes (50-100s), higher velocities ($\sim$100~km/s), are considerably taller (5-8~Mm) and rarely exhibit any recession (downfall) of material that is ejected upward. It has been demonstrated that both types of spicules undergo Alfv\'{e}nic motion \citep[][]{DePontieu2007b}. However, in terms of chromosphere/transition region connectivity, there are several questions that are inadequately explained at present, but may be critical in understanding the interface of the cool and hot solar atmospheres:\\
\indent 1.~What is the relationship between chromospheric spicules and the emitting transition region structures observed above the limb?\\
\indent 2.~How are these spicules heated and to what temperature?\\
\indent 3.~Is the transition region really composed of two physically different ``components'' and do these spicule types play a role?\\
\indent 4.~What mechanism drives Type-II spicules?\\
\indent 5.~Do the unresolved transverse and longitudinal motions of the spicules produce the observed non-thermal line widths?\\
These are points we wish to address with the WHI joint observations and the detailed work is in its early stages. The present paper discusses some of the interesting preliminary analysis pertaining to points 1 and 2.

\section{WHI JOP 204: Characterizing the Quiet Sun}   %%% Top level section head (remove "%" symbol)
The purpose of WHI JOP204\footnote{http://sohowww.nascom.nasa.gov/soc/JOPs/jop204.txt} was to characterize the behavior of the Quiet Sun; aiming to build a better picture of the relationship between chromospheric dynamic spicules, their extension into the transition region and (possibly) corona. In essence we hope to fully assess the role of magneto-convective driving mass and energy from the lower atmosphere into the quiet corona or solar wind. JOP~204 Observations took place between April 10th through April 16th, 2008 involving a host of observatories on the ground (NSO) and in space ({\em SOHO}, {\em TRACE}, {\em Hinode}, {\em STEREO}). We will discuss the preliminary analysis of data from {\em Hinode}/SOT, {\em TRACE} and {\em SOHO}/SUMER \citep[][]{Wilhelm1995} on April 12th 2008 on the South-West limb that ran from 14:00 - 17:00UT.

\begin{figure}
%\epsscale{0.75}
\plottwo{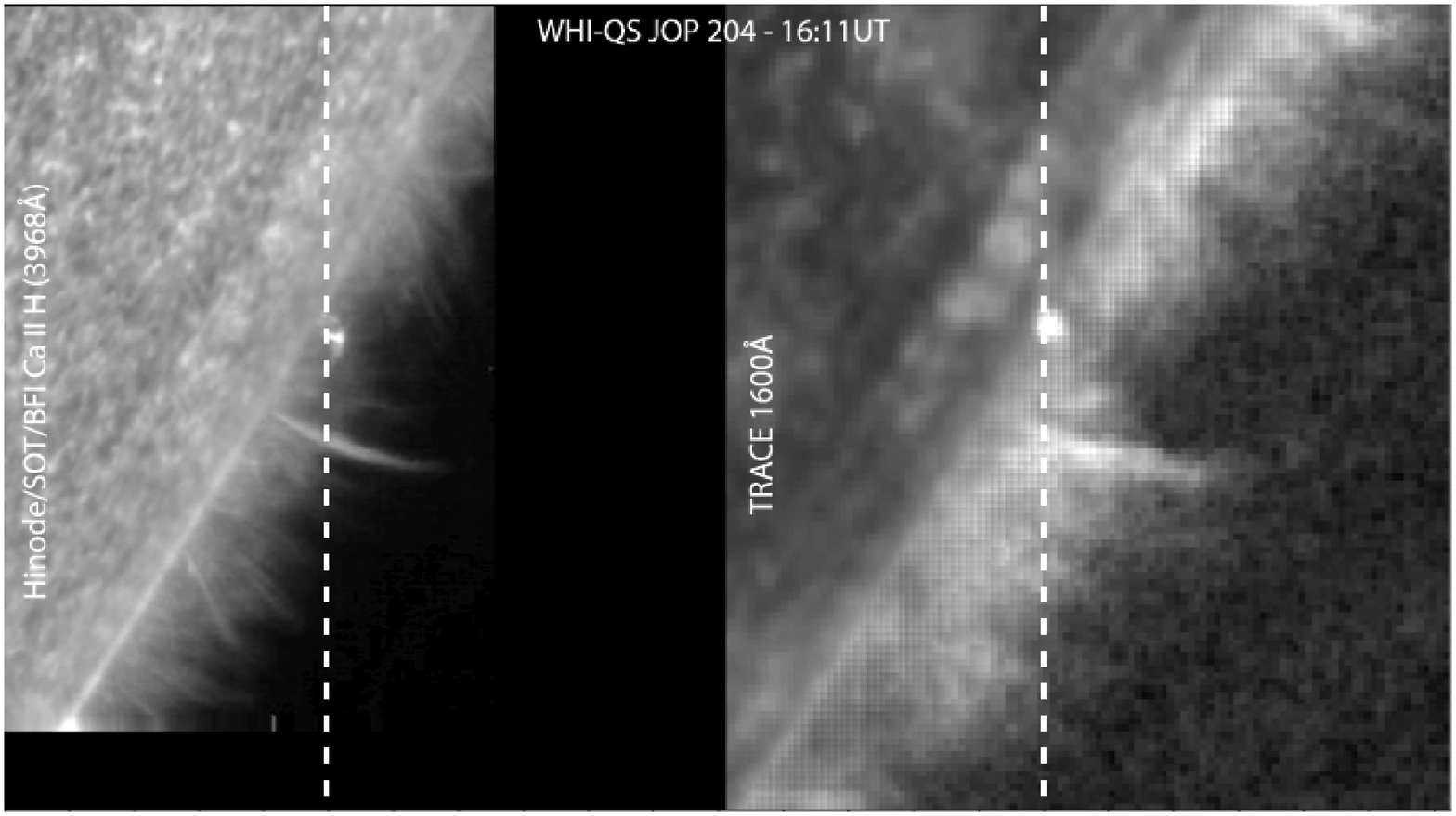}{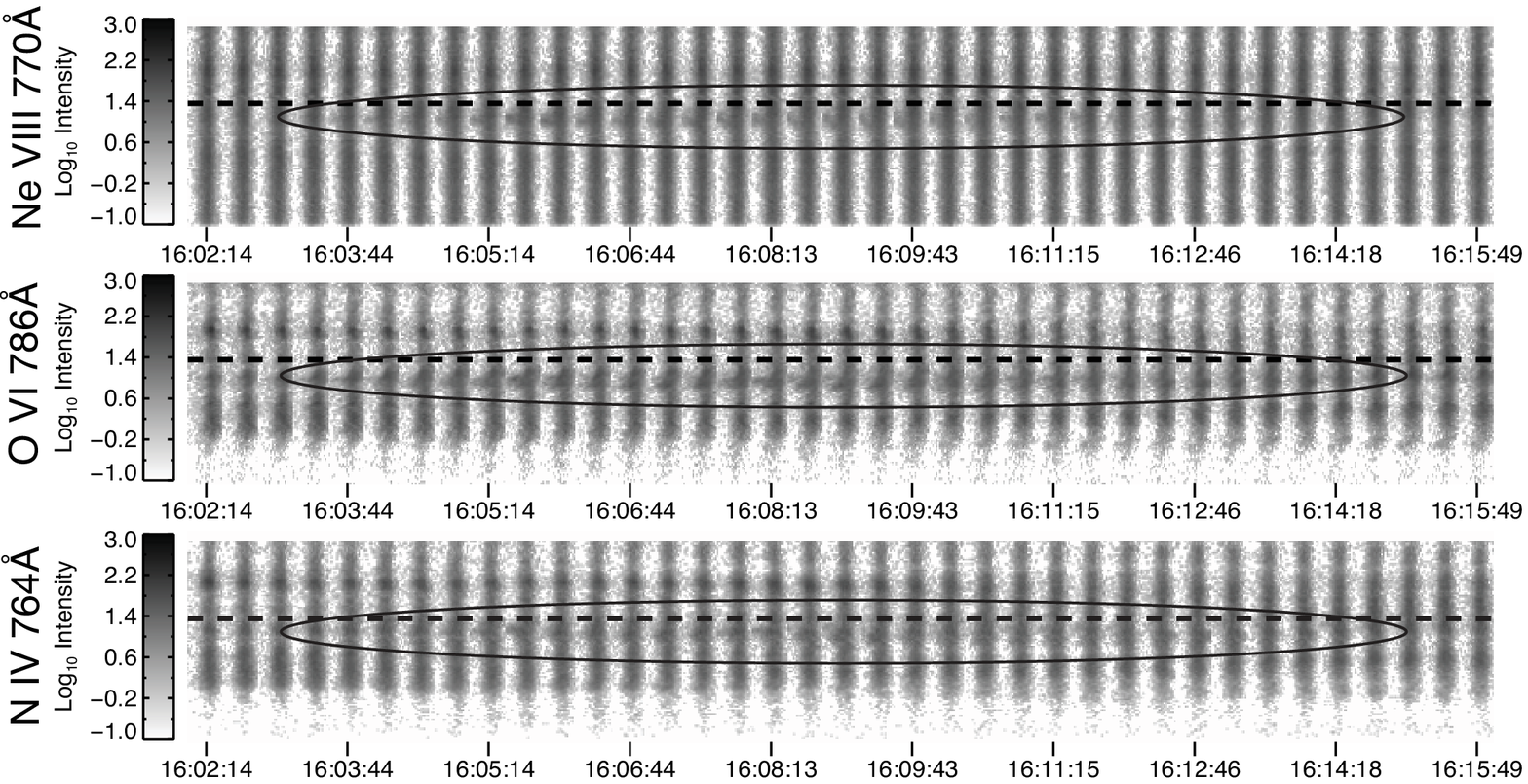}
\caption{Left: Comparison of the \hinode{} Ca~II~H and \trace{} 1600\AA{} passband images at 16:11UT in a 60\arcsec$\times$60\arcsec{} FOV centered on the middle portion of the SUMER slit (vertical dashed white line). This is the point in the image sequence where the macro-spicule is at its maximum extension. Right: Comparing consecutive SUMER spectral snapshots of N~IV 764\AA{} (bottom), O~VI 786\AA{} (middle) and Ne~VIII 770\AA{} (top) emission lines over the duration of the macro-spicule eruption (ellipse) with a reversed intensity scale. The horizontal dashed line denotes the position of the SUMER continuum limb. \label{f1}}
\end{figure}

The correspondence between above-the-limb structure in {\em Hinode}/SOT Ca~II~H and TRACE~1600\AA{} passband image sequences is striking - noting that the 1600\AA{} passband is dominated by 1548 and 1550\AA{} emission lines of C~IV there. Immediately, we see that while the emission observed in Ca~II is considerably finer in structure the gross behavior and dynamics present are repeated in the mid-transition region\footnote{Accompanying movies can be found at http://download.hao.ucar.edu/pub/mscott/Hinode2/.} indicating that material is at least heated to those temperatures as a spicule. By far the most easily identifiable event to transpire was the passage of a large macro-spicule across the SUMER slit at about 16:08UT, see Figure~\pref{f1}. Space-time analysis of the macro-spicule in the SOT and TRACE image sequences shows that the ejected material follows a parabolic trajectory (characteristic of Type-I spicules) with super-sonic launch and receding velocities of $\sim$40km/s seen some 7 minutes later. This event is clearly on a much larger spatial scale than the typical Type-I and II spicules which dominate the limb chromosphere and transition region. We suggest that the structure seen is a macro-spicule driven by the large-scale reconnection of the photospheric magnetic field on the supergranular scale \citep[e.g.,][]{McIntosh2007} with the material driven upward by a shock, given the parabolic spicule evolution observed

\begin{figure}
\plotone{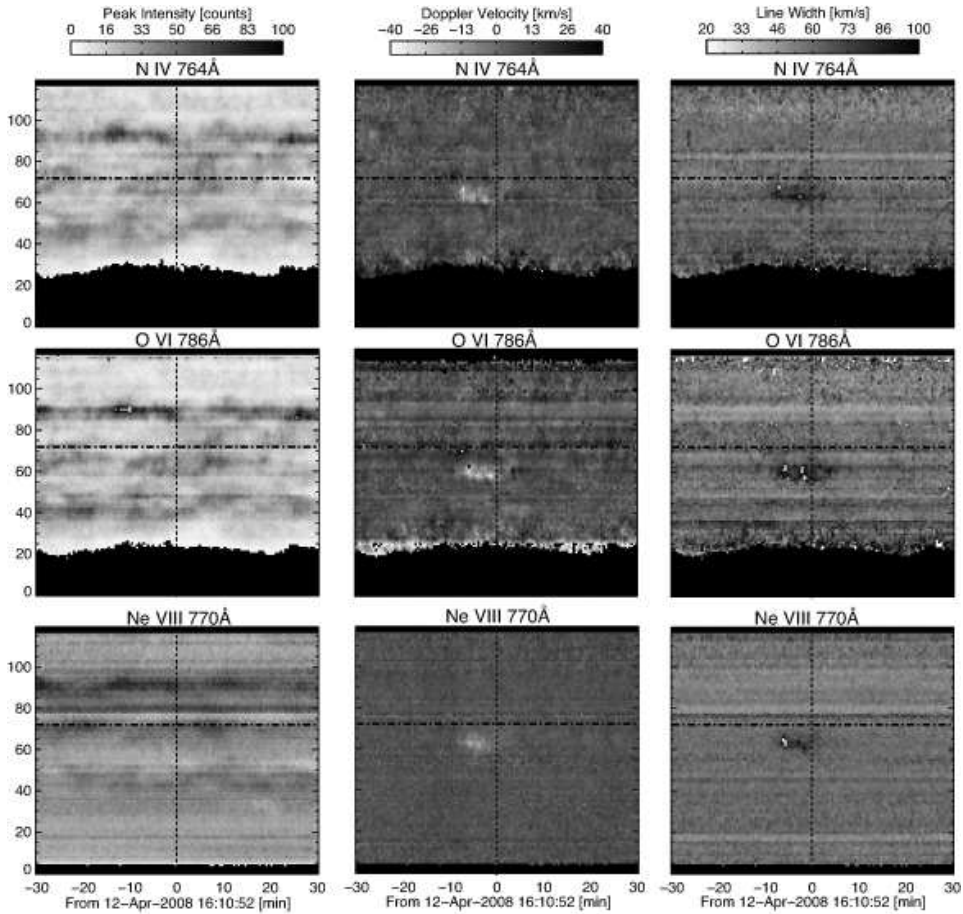}
\caption{From left to right SUMER Space-Time plots of line peak intensity, Doppler shift and 1/e line width determined from single Gaussian fits to the profiles N~IV 764\AA{} (top), O~VI 786\AA{} and Ne~VIII 770\AA{} emission lines around the time of the macro-spicule (t=0 is the dashed vertical line). The position of the continuum limb is shown as the dot-dashed horizontal lines. \label{f2}}
\end{figure}

\subsection{SUMER Multi-Wavelength Study of the Macro-spicule}
Studying the dynamic behavior of this event is made complex by the non-trivial evolution of the line profiles with the increasing formation temperatures of the lines as shown in Figs.~\pref{f1} and~\pref{f2}. The right side of Fig.~\pref{f1} shows spectral snapshots while Fig.~\pref{f2} shows the space-time evolution of the fitted (single Gaussian fit) spectra in the same lines. We see the upward phase of the event in {\em all} the emission lines as the line centers are blue-shift by $\sim$40km/s - consistent with the measurements from SOT - indeed the cooler emission (N~IV and O~VI) show parabolic behavior in the space-time plots. Further, we also see that the profiles are asymmetric and get significantly broadened ($\sim$60km/s, a non-thermal line width of $\sim$40km/s at these wavelengths - with SUMER detector B) over the entire duration of the event, indicative of unresolved velocities or active heating of the structure. Perhaps the most startling thing, and best seen in the bottom row of Fig.~\pref{f2}, is that the Ne~VIII emission of the macro-spicule is extremely weak (only a few of percent of the background Ne~VIII emission above the limb) but as the material starts to fall, and we start to see red-shift, the excess hot emission is {\em gone}; both the Doppler and line width signature of the macro-spicule disappears abruptly while the emission from the cooler lines show downward motion. The apparent disappearance of the hot emission provides a lot of questions for subsequent work and motivates the discussion below.

\section{Discussion}
We have been able to successfully co-align data from \hinode/SOT, \trace{} and \soho/SUMER for the WHI QS period on April 12 2008. The data provide a tantalizing look at the dynamic evolution of ejected transition material and its interaction with the corona which surrounds it \citep[][]{Judge2008}, the macro-spicule observed clearly reaches the formation temperature of Ne~VIII \citep[600,000K;][]{Mazzotta1998}. Further, we speculate that the hottest material seen, based on the lack of returning Ne~VIII emission, has actually become a constituent of the corona. It is probably not a coincidence that this emission line is formed above the temperature at which thermal conduction is critical in the energy balance of the plasma \citep[][]{Mariska1992} and that Ne~VIII is the coolest transition region species for which we see outflow in coronal holes \citep[e.g.,][]{McIntosh2007}. We suggest that the physical process required for such a ``hot mass-injection'' to the coronal magnetic system is intrinsically not one that can be studied in an single-fluid ideal MHD framework. Subsequent work will show that this is the proverbial tip of the iceberg in terms of our understanding of the transition region and its role in the heating and mass-loading of the outer solar atmosphere.

\acknowledgements %%% Text of acknowledgements runs on after this command.
\begin{small}
We would like to thank Alfred de Wijn (SOT), Werner Curdt (SUMER) and Dawn Myers (\trace) for conducting the observations discussed. The work presented here is supported by a grant NNX08AL22G to SWM and BDP. BDP is supported by the NASA \hinode{} grant to LMSAL (NNM07AA01C).The National Center for Atmospheric Research is sponsored by the National Science Foundation.
\end{small}


\begin{thebibliography}{}

%\bibitem[{{de Wijn} \& {De Pontieu}(2006)}]{deWijn2006}
%{de Wijn}, A.~G., \& De Pontieu, B., 2006, \aap, 460, 309

\bibitem[{{De Pontieu} {et~al.}(2007a)}]{DePontieu2007a}
{De Pontieu}, B., et~al., 2007a, \apj, 655, 624

\bibitem[{{De Pontieu} {et~al.}(2007b)}]{DePontieu2007b}
 De Pontieu, B., et~al., 2007b, Science, 318, 1574
 
\bibitem[{{De Pontieu} {et~al.}(2007c)}]{DePontieu2007c}
 {De Pontieu}, B., {et~al.}, 2007c, Pub. Ast. Soc. Jap., 59, 655
 
 \bibitem[{{Hansteen} {et~al.}(2006)}]{Hansteen2006}
{Hansteen}, V.~H., et~al. 2006, \apj, 647, L73

 \bibitem[{{Judge}(2008)}]{Judge2008}
Judge, P.~G., 2008, \apj, 683, L87

 \bibitem[{{Kosugi} {et~al.}(2007)}]{Hinode}
Kosugi, T., et~al., 2007, \solphys{}, 243, 3

\bibitem[{{Mariska}(1992)}]{Mariska1992}
Mariska, J.~T., 1992, ``The Solar Transition Region", Cambridge University Press

\bibitem[{{Mazzotta} {et al.}(1998)}]{Mazzotta1998}
Mazzotta, P., Mazzitelli, G., Colafrancesco, S., \& Vittorio, N. 1998, \aaps, 133, 403

\bibitem[{{McIntosh} {et~al.}(2007)}]{McIntosh2007} 
McIntosh, S.~W., et al. 2007, \apj, 654, 650

\bibitem[{{Tsuneta} {et~al.}(2008)}]{Tsuneta2008}
Tsuneta, S., et~al., 2008, \solphys, 249, 167
 
\bibitem[{{Wilhelm} {et~al.}(1995)}]{Wilhelm1995}
Wilhelm, K., et~al., 1995, \solphys, 162, 189

\end{thebibliography}
\end{document}